# Unzipping and binding of small interfering RNA with single walled Carbon Nanotube: a platform for small interfering RNA delivery

*Mogurampelly Santosh[1,3], Swati Panigrahi[2], Dhananjay Bhattacharyya[2], A. K. Sood[3] and Prabal K Maiti [*1,3]*

[1] Centre for Condensed Matter Theory, Indian Institute of Science, Bangalore 560012, India.

[2] Biophysics Division, Saha Institute of Nuclear Physics, Kolkata 700064

[3] Department of Physics, Indian Institute of Science, Bangalore 560012, India.

**Abstract**

In an effort to design efficient platform for siRNA delivery, we combine all atom classical and quantum simulations to study the binding of small interfering RNA (siRNA) by pristine single wall carbon nanotube (SWCNT). Our results show that siRNA strongly binds to SWCNT surface via unzipping its base-pairs and the propensity of unzipping increases with the increase in the diameter of the SWCNTs. The unzipping and subsequent wrapping events are initiated and driven by van der Waals interactions between the aromatic rings of siRNA nucleobases and the SWCNT surface. However, MD simulations of double strand DNA (dsDNA) of the same sequence show that the dsDNA undergoes much less unzipping and wrapping on the SWCNT in the simulation time scale of 70 ns. This interesting difference is due to smaller interaction energy of thymidine of dsDNA with the SWCNT compared to that of uridine of siRNA, as calculated by dispersion corrected density functional theory (DFT) methods. After the optimal binding of siRNA to SWCNT, the complex is very stable which serves as one of the major mechanisms of siRNA delivery for biomedical applications. Since siRNA has to undergo unwinding process with the effect of RNA-induced silencing complex, our proposed delivery mechanism by SWCNT possesses potential advantages in achieving RNA interference (RNAi).

Keywords: siRNA, CNT, Unzipping, Wrapping, van der Waals Interaction and MD Simulations.

---

[*]Correspoding Author: maiti@physics.iisc.ernet.in

# 1. Introduction

RNA interference (RNAi) is a powerful technology for controlling the expression of genes in biomedical applications. Small interfering ribonucleic acid (siRNA) molecules (typically 21 to 23 nucleotides) are being actively studied due to their potential influence on cell functionality and applications in medicine [1,2]. RNAi is a cellular process in which a double stranded RNA (dsRNA) reduces a specific gene expression. The base-pairing of siRNA with messenger RNA (mRNA) sequence silences the encoded protein. The mechanism of RNAi involves RNA-induced silencing complex (RISC) comprising of Dicer, Argonaute2 and siRNA binding protein that induces unzipping of siRNA into two single strand RNAs [3-7]. One of these two strands acts as a guiding strand to specifically base-pair with mRNA. Generally for efficient gene silencing, chemically unmodified siRNAs are rapidly degraded in serum and hence siRNAs have to be bound with transfecting carriers [8]. We show the unwinding enhanced siRNA binding to carbon nanotube (CNT) and propose siRNA delivery to target cell for achieving RNAi without degradation. Possible transfecting carriers are linear or branched cationic polymers (dendrimers) [9-13], cationic lipids [14,15], carbon nanotubes (CNTs) [16-19], cell penetrating peptides [20,21] and few proteins [22,23]. Though siRNA molecules are proved to be potential silencers of gene expression that can have extraordinary treatment capabilities of HIV, hepatitis and cancer [24-30], efficient delivery of these molecules to the target cell is a big challenge today.

In this work, we address the question of using SWCNT as efficient carrier. CNTs functionalized with polymers such as Polyethylene glycol (PEG), CONH-(CH2)(6)-NH3+Cl- or single stranded DNA (ssDNA) are efficient transporters of siRNA into human T cells and primary cells [17,19,31-33]. However, the structural and energetic changes taking place in siRNA while binding to the SWCNT surface are not known. In comparison, wrapping of single stranded DNA (ssDNA) [34-37] and double stranded DNA (dsDNA) [38] on the SWCNT surface have been studied recently. With this in view, we have studied the thermodynamics and conformational properties of siRNA-SWCNT complex through all atom molecular dynamics (MD) simulations and ab-initio quantum calculations.

# 2. Computational details
## 2.1 Molecular Dynamics Simulations

Molecular dynamics simulations were carried out using AMBER9 suite of programs [39] using the AMBER 2003 (along with ff99) [40] force fields and the TIP3P model [41] for water. Latest improvements of torsion angle parameters are reported for RNA in a new force field called parmbsc0 and its variants [42,43]. However we have used ff99 for siRNA instead of parmbsc0 since the conformational changes of siRNA reported here are not artefacts of the wrong torsion angle parameterization in ff99. Further, we intend to compare the present results with our previous simulations on siRNA that used ff99 [11-13]. Notwithstanding this, we have also performed a few simulations with parmbsc0 and obtained similar conformational changes as reported here which are shown in the supplementary Fig. S1 and S2 [44] and, therefore, conclude that the ff99 parameters are reliable (for example see [45-47]). The initial structure of the siRNA was taken from the protein data bank (PDB code: 2F8S) [48]. The sequence of the siRNA used is UUr(AGA CAG CAU AUA UGC UGU CU)$_2$UU with sticky ends of sequence UU on the two ends of the strands. We have built armchair SWCNT of various diameters; (5, 5) ($d$ = 6.68 Å), (6, 6) ($d$ = 8.02 Å), (7, 7) ($d$ = 9.36 Å) and (8, 8) ($d$ = 10.69 Å). We choose carbon nanotubes to be 142 Å in length to ensure sufficient sliding length for the siRNA before stable binding. In the initial configuration, the siRNA is placed on the nanotube surface such that the nanotube axis and siRNA helix axis are nearly parallel to each other. The siRNA-SWCNT complexes were then solvated separately with TIP3P water box [41] of dimensions as shown in Table 1 using the LEaP module in AMBER9. The box dimensions were chosen such that there is at least 20 Å solvation shell in all directions from the surface of siRNA-CNT complex during the entire simulation. In addition, some water residues were replaced by 44 $Na^+$ counterions to neutralize the negative charge on the phosphate backbone groups of the siRNA structure. The initial system containing siRNA-(6, 6) CNT with added water and neutralizing counterions is shown in Fig. 1. In separate simulation runs, additional NaCl residues were added to prepare system at 10 *mM* and 150 *mM* salt concentrations. These three salt concentrations (0 *mM*, 10 *mM* and 150 *mM*) were studied in view of the importance of electrostatic screening in binding mechanisms of siRNA with the SWCNT surface. For comparison, we have also simulated dsDNA adsorption on (5, 5) and (6, 6) CNTs of the same length at 300 K. The dsDNA has the same length and sequence as the siRNA where the nucleobase Uracil (U) is replaced by nucleobase Thymine (T), i.e., TTd(AGA CAG CAT ATA TGC TGT CT)$_2$TT. In another simulation, we choose a random sequence of dsDNA as given by d(GCA TGA AAT GCT TAA AGC TTA C)$_2$. The full details of the various systems studied, number of NaCl residues, water residues, box dimensions and total number of atoms are summarized in Table 1.

For simulating the CNT, carbon atoms are modelled as uncharged Lennard-Jones particles with sp$^2$ hybridization according to the parameters from AMBER03 force field (type CA). In addition, bonded interactions viz., stretching, torsion and dihedral terms were also included. We have used the same force field for the CNTs earlier in the context of water transport through them [49]. To keep the CNT fixed during simulations, all the atom positions were constrained with harmonic potential of spring constant of 1000 *kcal/mol-Å$^2$*. The translational centre of mass motions were removed every 1000 steps. The long range electrostatic interactions were calculated with the Particle Mesh Ewald (PME) method [50] using a cubic B-spline interpolation of order 4 and a $10^{-5}$ tolerance set for the direct space sum cutoff. A real space cutoff of 9 Å was used both for the long range electrostatic and short range van der Waals interaction with a non-bond list update frequency of 10. We have used periodic boundary conditions in all three directions and the bond lengths involving bonds to hydrogen atoms were constrained using SHAKE algorithm [51]. This constraint enabled us to use a time step of 2 fs for obtaining a long trajectory. During the minimization, the siRNA-SWCNT complex structures were fixed in their starting conformations using harmonic constraints with a force constant of 500 *kcal/mol-Å$^2$*. This allowed the water molecules to reorganize which eliminates bad contacts with the siRNA and the CNT structures. The minimized structures were then subjected to 40 ps of MD, using 1 fs time step for integration. During the constant volume - constant temperature (NVT) MD, the system was gradually heated from 0 to 300 K using weak harmonic restrains of 20 *kcal/mol-Å$^2$* on the solute to its starting structure. This allows slow relaxation of the siRNA-CNT structure. Subsequently, simulations were performed under constant pressure - constant temperature conditions (NPT), with temperature regulation achieved using the Berendsen weak coupling method [52] (0.5 ps time constant for heat bath coupling and 0.5 ps pressure relaxation time). NPT-MD was used to get the correct (experimental) solvent density. Finally, for analysis of structures and properties, we have carried out 50-100 ns of NVT MD with 2 fs integration time step using a heat bath coupling time constant of 1 ps. The trajectories were saved at a frequency of 2 ps.

**2.2 Quantum Chemical Calculations**

We have also carried out quantum chemical analysis with dispersion correction (DFT-D) to understand the interactions of the CNT with the DNA and RNA. The valencies of the carbon atoms at the ends of the (6, 6) CNT were satisfied by adding necessary hydrogen atoms. The structural features that distinguish RNA from DNA are the presence of uracil base and 2'-OH groups of the ribose sugars. The four systems modeled by us are as follows: (i) (6, 6) CNT with one uracil nucleobase, (ii) (6, 6) CNT with one thymine nucleobase, (iii) (6, 6) CNT with one uridine nucleoside (uracil attached with C3'-endo ribose sugar) and (iv) (6, 6) CNT with thymidine nucleoside (thymine nucleobase attached with C2'-endo deoxyribose sugar) as shown in Fig. S3(a). These initial structures were built using MOLDEN [53] software. Initially the nucleobases were placed in parallel to the CNT surface. Free geometry optimization of all the four systems discussed above were carried out using Gaussian09 [54] without any constraints using density functional theory with WB97XD/6-31g(d,p) basis set [55], which includes dispersion correction, giving rise to energy $E_{XY}$ (potential energy of CNT + nucleobase or nucleoside). We have also optimized the isolated CNT, giving rise to energies $E_{X0}$, the two nucleobases and the two nucleosides in un-complexed isolated form having energy $E_{Y0}$. Optimizations of such supramolecular complex systems are associated with a Basis Set Superposition Error (BSSE), which mainly arise due to overlapping of the optimized orbitals. This error has been corrected by Boys-Bernardi function counterpoise method [56]. The BSSE corrected interaction energies ($E_{int}$) of each system were calculated using $E_{int.} = E_{XY} - E_{X0} - E_{Y0} + BSSE$. We have also carried out the Mulliken charge analysis of all the optimized systems to analyze the charge transfer taking place, if any, between the CNT and the associated nucleobase/nucleoside.

## 3. Results and discussion
### 3.1 Unzipping and wrapping of siRNA on binding to (6, 6) CNT

The snapshots of the siRNA and (6, 6) CNT complex at 0 ns, 15 ns, 30 ns and 45 ns shown in Fig. 2 (a) and (b) correspond to horizontal and vertical views of the complex with respect to the CNT axis, respectively. The horizontal view shows unzipping of the base-pair at various instants of time and the vertical view presents the wrapping of siRNA around the CNT surface. The unzipping and wrapping of a few base-pairs (~ 6 to 7) at end **A** of the siRNA was observed after about 12 ns as shown in Fig. 2. Quite intuitively at the other end **B** of siRNA also, 2 to 3 base-pairs were unzipped and wrapped around the CNT. This wrapping at end **B** provides a constraint for further unzipping of siRNA from the side **A**. In order to further unzip at both the ends, the torsional relaxation has to take place which happens only when the unzipping force is more than the binding force. Since this will not be achieved without an external force, the complex is very stable over the rest of entire MD simulation time of 70 nanoseconds. We note that the wrapping/binding occurs without any CNT surface chemistry or functionalization. To our knowledge, this is the first demonstration of the unzipping and wrapping of siRNA around the CNT.

To understand the thermodynamics of the binding events, we have calculated van der Waals interaction between the siRNA and the (6, 6) CNT surface at various time intervals, shown only for 5 ns and 15 ns in Fig. 3. Different contributions to the total energy of the siRNA such as electrostatic interaction, van der Waals interaction and bonded interaction energies were analyzed. We find that the van der Waals interaction is responsible for the observed unzipping and wrapping of the siRNA. Other contributions to the total energy do not contribute much to the binding as a function of time. Entropy calculations support this inference as will be discussed in section 3.2. In Fig. 3, $r = 0$ corresponds to the centre of mass of siRNA as shown in the schematic diagram. The interaction coordinate $r$ was projected along the nanotube axis $\hat{n}$. The van der Waals interaction energy between aromatic rings of siRNA and CNT surface ($-\phi(r)$) is more symmetrical with respect to $r = 0$ at 5 ns than at 15 ns, with more tendency of binding of one end of the siRNA strands to the CNT surface. This asymmetry in the interaction can result in the motion of siRNA on the nanotube (seen in Fig. 3) until it finds its optimal binding position. Now we address how the surface area of nanotube does affects the unzipping process and subsequent wrapping of the siRNA on the nanotube by calculating the binding free energy of siRNA with nanotubes of various diameters.

### 3.2 Binding free energy of siRNA on CNT surface

The binding free energy for the non-covalent association of two molecules in solution can be written as $\Delta G(A + B \rightarrow AB) = G_{AB} - G_A - G_B$. Accordingly,

$$\Delta G_{bind} = \Delta H_{bind} - T\Delta S_{bind} \qquad (1)$$

where $\Delta H_{bind}$ is the change in enthalpy and is calculated by summing the gas-phase energies ($\Delta E_{gas}$) and solvation free energies ($\Delta G_{sol}$). Note $E_{gas} = E_{ele} + E_{vdw} + E_{int}$, where, $E_{ele}$ is the electrostatic energy calculated from the Coulomb potential, $E_{vdw}$ is the non-bonded van der Waals energy and $E_{int}$ is the internal energy contribution arising from bonds and torsions. Further, $G_{sol} = G_{es} + G_{nes}$, where $G_{es}$ is the electrostatic energy calculated from Generalized Born (GB) method [57-59] and $G_{nes}$ is the non-electrostatic energy calculated as $\gamma \times SASA + \beta$ where $\gamma$ is the surface tension parameter ($\gamma = 0.0072$ kcal/mol-Å$^2$; $\beta = 0$ kcal/mol) and $SASA$ is the solvent-accessible surface area of the molecule. All these enthalpy calculations were done using MM-GBSA module of AMBER [39] suite of programs. The entropy is calculated using two-phase thermodynamic (2PT) model developed by Lin *et al* [60], based on density of states (DoS) function. The DoS function can be calculated from the Fourier transform of the velocity autocorrelation function which provides information on the normal mode distribution of the system, with the zero frequency intensity in DoS corresponding to the diffusivity of the system [61]. The 2PT method was successfully used to estimate the entropy and energetics of molecular fluids [60,62] from the trajectory of molecular dynamic simulations.

The trajectory for enthalpy ($\Delta H_{bind}$) calculation was chosen such that the binding energy is minimum and stable over at least 10 ns which requires extensive MD simulations (60-80 ns in this study) from a suitable initial configuration of siRNA-CNT complex. In Fig. 4(a), we plot enthalpy contribution to the total binding free energy as a function of time for siRNA on (6, 6) CNT at 300

K. In the plot, arrow marked region is the most optimum bound state for siRNA on (6, 6) CNT which is constant for over 10 ns with fluctuations ranging 5 % of its average in optimum bound state. From the stable trajectory of siRNA-CNT complex, the enthalpy contribution ($\Delta H_{bind}$) to the total binding free energy was calculated for 250 snapshots separated each by 2 ps. For calculating entropy at the optimum bound state where the minimum value of enthalpy is seen, we simulate the system for 20 ps with the trajectory saved at a frequency of 4 fs. Entropy calculations were done for 10 such successive sets of each 20 ps MD trajectories with velocity and coordinates saved at every 4 fs. For all these 10 successive sets, the velocity autocorrelation function is seen to converge in less than a correlation time of 10 ps. The enthalpy and entropy are calculated for the siRNA-CNT complex and individual siRNA and CNT separately. The siRNA is simulated for 20 ns without the CNT. The entropy of siRNA when complexed with CNT increases with the tube diameter due to more unzipping of the base-pairs in siRNA leading to large microstates available for siRNA. However the contribution of entropy ($-T\Delta S_{bind}$) to the total binding free energy is very less compared to the enthalpy contribution. The enthalpy ($\Delta H_{bind}$) and entropy contribution ($-T\Delta S_{bind}$) in Eq. (1) gives the total binding free energy ($\Delta G_{bind}$) of siRNA. Fig. 4(b) shows $\Delta G_{bind}$ (at 300 K) as a function of tube diameter revealing that the binding energy increases with the diameter of the nanotube. This is easy to rationalize because siRNA encounters more surface area with increased diameter of the tube. The binding of siRNA increases as the nanotube radius of curvature increases and is maximum when the radius of curvature of the nanotube is equal to the radius of siRNA. The binding of siRNA to the CNT involves unzipping of stacked base-pairs which has its own characteristic length scale to match the length scale of the CNT curvature. The binding free energy of siRNA on (5, 5) CNT is -189.0 ± 15.3 *kcal/mol* which increases to -301.0 ± 11.4 *kcal/mol* for (8, 8) CNT. Snapshots of siRNA on CNT for various diameters in the most optimum bound configuration are shown in Fig. 5 in horizontal and vertical view with respect to CNT axis. All the above studies are for NaCl concentration of 0 *mM*. Physiological NaCl concentrations are of 100 to 150 *mM* in cell and can affect the binding mechanism of siRNA-CNT complex during delivery. We have, therefore studied siRNA binding mechanism to (6, 6) CNT at 300 K by increasing NaCl concentration. For (6, 6) CNT at 300 K, the binding free energy decreases from -230.0 ± 4.8 *kcal/mol* to -145.7 ± 5.5 *kcal/mol* when NaCl concentration increased to 150 *mM*. With more added salt in the solvent, the electrostatic screening increases resulting in reduced phosphate-phosphate electrostatic repulsion in the backbone of siRNA. Therefore the stretch modulus of siRNA increases giving higher stability to siRNA with increasing salt concentration [63-65] which reduce the propensity of siRNA unzipping. This leads to lesser binding affinity of siRNA with CNT compared to the case of lower salt concentration. The electrostatic screening due to ionic charges at 150 $mM$ NaCl concentration makes less efficient binding of the siRNA with the CNT. If the wrapping and binding are governed by electrostatic interaction, $Na^+$ counterions should strip away from the siRNA-(6, 6)CNT complex which can result in increase in the entropy of the counterion. However, our entropy calculation shows that $Na^+$ counterions do not gain entropy for the 0 mM NaCl run. This could be due to the fact that the Na+ ions are not getting stripped away from the complex (see table 2) as can be seen from the radial distribution function of $Na^+$ counterions from phosphate groups of the siRNA backbone (see Fig. S4). This also demonstrates that van der Waals interactions drive the wrapping and binding of siRNA to the CNT surface. However, for higher salt concentrations, $Na^+$ ions do gain entropy when NaCl concentration increases because of ion pairing. In Fig. S5, we give the power spectrum of translational, rotational and vibrational entropy $s(v)$ of siRNA on binding to the CNT at 15 ns. In all the CNT of various diameters as discussed above, the bound complex is very stable with binding free energy fluctuating within 5-10 % after optimal binding. This stable complex structure with some base-pairs already unzipped in siRNA can be delivered to the target virus infected cell for RNAi applications. Since siRNA has to undergo unwinding process with the effect of RISC, our proposed delivery mechanism by CNT possesses potential advantages in achieving RNAi. Toxic effects of CNT inside cell can be suppressed with proper surface functionalization [66-68]. Functionalization of CNT with dendrimer and its complexation with siRNA study is under progress. Such study can address issues such as binding of siRNA to functionalized CNT and the effect of functionalization on toxicity also.

**3.3 Watson-Crick (WC) H-bonds**

In many biological phenomena where nucleic acids are involved, Watson-Crick H-bonds manifest the underlying deformation mechanism of nucleic acid molecule [69]. To demonstrate the unzipping of siRNA when adsorbed to nanotube, we have calculated the number of intact Watson-Crick H-bonds in siRNA with time for all the CNT diameters studied. We have used geometry based measurement criteria for the H-bond calculation. The H-bond is represented as D-H…A; where D is the donor and A is the acceptor which is bonded to D through H atom, three dots denote H-bond and hyphen denotes a covalent bond. In case of siRNA or DNA, D is nitrogen (N) atom and A is either N or oxygen (O) atom depending on the A-U (A-T) or G-C base-pairing. When the distance between D and A atoms is less than 2.7 Å and the angle ∠DHA is greater than 130˚, the atom A is H-bonded to atom D, otherwise the H-bond is broken. Fig. 6 shows the intact H-bonds as a function of time. In siRNA, there are 48 maximum intact H-bonds. It is known that at room temperature a small fraction of transient broken H-bonds can exist in siRNA/DNA due to thermal fluctuations. In our simulation we find the maximum of 47 intact H-bonds for all diameters at t = 0. Fig. 6 shows that the number of WC H-bonds decreases with time since the siRNA base-pairs get unzipped. The unzipping is more when the diameter is increased. At long times, the number of intact WC H-bonds decreases with increasing CNT diameter. There are as low as 22 intact WC H-bonds in siRNA when bound to (8, 8) CNT. Since the number of intact WC H-bonds decrease with tube diameter, the unzipped base-pairs are then free to bind to the nanotube surface which, in turn, increases the binding energy of the siRNA. Thermal melting of WC H-bonds makes siRNA less stable and the CNT acts as a supporting substrate to bind with large binding energy of -232.5 ± 4.9 *kcal/mol* at 300 K. As the WC H-bond breaking increases, the entropy of siRNA also increases significantly. It would be interesting to study the kinetics of the unzipping and adsorption of individual nucleobases when binding to CNT. The structural aspects including number of closest atoms of siRNA to nanotube surface and deformation mechanism are discussed in section **3.4**.

**3.4 Structural deformation**

Snapshots shown in Fig. 2 and Fig. 5 indicate that the siRNA molecule exhibit large structural deformation on binding to the carbon nanotube. This structural deformation is characterized by the number of siRNA atoms that come close to the nanotube in a specified cutoff distance and root mean square deviation (RMSD) of siRNA with respect to its crystal structure. We calculate the number of close contacts of siRNA within 5 Å from the surface of CNT as follows:

$$N_c(t) = \sum_{i=1}^{N_{CNT}} \sum_{j=1}^{N_{siRNA}} \int_{r_i}^{r_i + 5 A^\circ} \delta(r(t) - r_j(t)) dr \qquad (2)$$

Here $N_{CNT}$ and $N_{siRNA}$ are the total number of atoms in CNT and siRNA, respectively, and $r_j$ is the distance of $j^{th}$ atom of siRNA from $i^{th}$ atom of CNT. We plot the number of close contacts ($N_c$) in Fig. 7 for various nanotube diameters. For all the nanotube diameters (Fig. 7), the number of average close contacts of siRNA gradually increases as siRNA wraps around the nanotube. The maximum value of $N_c$ increase with increasing the nanotube diameter, 345 for (5, 5) and 508 for (8, 8). After unzipping of a few base-pairs, they come closer to the CNT surface in order to wrap around it, resulting in higher value of $N_c$. As discussed previously, Fig. 3 signifies such van der Waals interaction strength and its impact on siRNA binding. At 10 *mM* and 150 *mM* of NaCl for (6, 6) nanotube at 300 K, the maximum of average number of close contacts is less than that of zero salt case. The lower binding affinity at 10 *mM* and 150 *mM* NaCl concentration is due to the screening of ionic charges which results in conformational changes of the siRNA. For higher salt case, we expect the unzipping and wrapping of the siRNA in the same manner as in the case of 0 *mM* but the process takes long time. At present, for 150 *mM* case, simulations up to 50 ns brings out 4-5 base-pairs unzipped. Hence it is clear that the wrapping and binding of siRNA on CNT surface depends on NaCl concentration. Since most of the physiological conditions correspond to 100-150 *mM* NaCl concentration, the longer time taken by the siRNA to bind to the CNT should not be an obstacle for medical applications.

Another important property that quantifies the amount of structural deformation in siRNA as it binds to CNT is root mean square deviation (RMSD) calculated as

$$RMSD = \sqrt{\frac{\sum_{i=1}^{N_{siRNA}} \left( \vec{r}_i - \vec{r}_{i,crystal} \right)^2}{N_{siRNA}}} \quad (3)$$

Where $\vec{r}_{i,crystal}$ is the $i^{th}$ atom's position vector of reference crystal structure of siRNA that was taken after initial minimization. The siRNA has translational and rotational motion with respect to the nanotube axis $\hat{n}$. Initially RMSD is ~ 2 Å in all the CNT cases due to the structural deformation in siRNA caused during 120 ps of NPT simulation stage. RMSD increases with time and saturates to 7-12 Å depending on the diameter of the CNT. In the most optimum bound configuration, the average RMSD ($\langle RMSD \rangle_{bound}$) of siRNA is 9.02 ± 0.30 Å, 11.21 ± 0.62 Å, 11.78 ± 0.26 Å and 11.98 ± 0.22 Å for (5, 5), (6, 6), (7, 7) and (8, 8) CNTs, respectively. As the binding is increasing with CNT diameter, the structure deforms largely leading to an increase in $\langle RMSD \rangle_{bound}$. In the long time stable bound configuration, RMSD fluctuates with standard deviation of 0.6 Å indicating the intrinsic dynamic nature of siRNA binding to CNT.

### 3.5 Comparing adsorption of siRNA and dsDNA

Adsorption of dsDNA with and without sticky-ends on (6, 6) CNT at 300 K was also studied for a qualitative comparison with the siRNA adsorption. The sequence of dsDNA with sticky-ends is same as siRNA sequence except thymine is in place of uracil. dsDNA without sticky-ends has a random sequence. In both the cases, dsDNA gets adsorbed to CNT with insignificant unzipping in the long simulation time of 70 ns. This can be explained due to the relatively strong A-T base-pair interaction energy compared A-U base-pair interaction energy [70]. Our results of dsDNA adsorption without sticky-ends on CNT are in excellent agreement with an earlier study [38] where only adsorption was reported. The van der Waals attraction, binding energy, WC H-bonds (Fig. 6) and the number of close contacts (Fig. 7) of dsDNA on the CNT are less compared to that of siRNA. However there is more fluctuation in the RMSD due to lower binding to CNT. The snapshots of dsDNA with the sticky-ends adsorbed on (6, 6) CNT at 300 K at various instants of time are shown in Fig. S6. To understand the role of sticky-ends of siRNA in unzipping and wrapping process, we have performed a simulation of RNA with the same sequence as siRNA but without any sticky-ends. Interestingly, RNA unzipping and wrapping around CNT is very less compared to the case where sticky-ends are present in siRNA. The RNA stays adsorbed on the CNT surface with linear translational motion along the CNT axis during the entire 50 ns long simulation. Therefore, the sticky-ends enhance the unzipping and wrapping of siRNA on CNT. On the other hand, sticky-ends do not help dsDNA in unzipping due to relatively stronger A-T base-pair energy compared to the A-U base-pair energy. Snapshots of RNA without sticky-ends adsorbed on nanotube are shown in Fig. S7.

**3.6 siRNA vs dsDNA: insights from Quantum mechanical calculations**

We find interaction energy of the CNT-uracil nucleobase complex is -9.64 *kcal/mol* whereas that of the CNT-thymine nucleobase complex is about -12.25 *kcal/mol*. Interaction energy between thymine and CNT was calculated earlier using Hartree-Fock and related methods giving significant attraction between the two [71]. The energies using more robust density functional theory which includes dispersion correction (DFT-D), the value of interaction energy of thymine obtained in our method are quite similar to that of the previous estimate (-11.3 *kcal/mol*), without dispersion and electron correlation effects. The difference in energy between thymine and uracil bases obtained in our method may arise due to the stronger non-polar interaction between the methyl groups of thymine and the carbon atoms of the CNT. Since sugar backbone plays a crucial role in maintaining the structure and stability of RNA or DNA, we have also optimized the CNT-uridine and CNT-thymidine complexes including sugars attached to the nucleobases. This now represents a more accurate model of CNT-siRNA complex. The interaction energy values shown in Table 3 indicate that CNT with uridine has stronger binding (-18.72 *kcal/mol*), than that of the CNT with thymidine (-16.25 *kcal/mol*). This is due to possibility of weak hydrogen bond formation between the three -OH groups of uridine molecule with carbon atoms of CNT. We believe this favours the siRNA unzipping and subsequent wrapping on CNT whereas only adsorption of dsDNA on CNT is observed. The uridine molecule forms two H-bonds involving O5'-H5' of ribose sugar with two carbon atoms at edge-1 (Fig. S3 (a(iii))) of CNT with H-bond distances of 2.62 Å and 2.52 Å and associated O-H...C angles of 150.81°  and 166.16°, respectively. Another H-bond is found between the O3'-H3' of the ribose sugar ring with one of the carbon atom that lies towards the middle of the CNT with H-bond distance and angle of 2.50 Å and 154.39°, respectively. Therefore, uridine is able to form three good H-bonds with the CNT and stabilize the system provided the carbon atoms have sufficient negative charges.

We have analysed partial charge of all the atoms of CNT calculated by Mulliken population analysis. As expected the terminal C-H groups are slightly polar, the carbons not bonded to hydrogen are neutral (carbons of C-C group, those lies inwardly at the terminal edges, Fig. S3 (b)) and the central carbons present at the middle region of the CNT have nearly zero charge. Moreover all the charges at the two edges are symmetrically distributed with zero standard deviations when it is not complexed with any nucleobase/nucleoside. In presence of thymine or uracil bases, the properties of the CNT remain nearly unchanged. A presence of nucleoside residues, particularly the uridine nucleoside, breaks the symmetry of the CNT significantly (shown in Table 4), as detected from the larger standard deviations of the charges of different groups of atoms. We have classified the CNT atoms into the following types (Fig. S3 (b)): i) carbon atoms at the edges which are not bonded to any hydrogen C-C carbons, ii) carbon atoms of C-H group and iii) carbon atoms which lie in the central region of CNT. The terminal atoms can be further classified into edge-1 and edge-2, depending on proximity to the binding nucleoside. When uridine binds to CNT, partial charges of the C-C carbon atoms at the edge-1 change significantly. In this case, the average charge of the carbon atom decreases and the standard deviation increases, which signifies delocalisation of the charges. In presence of the polar uridine, the electrons of the CNT move significantly through the extended conjugation and accumulate near the uridine. The thymidine also alters charges of these carbon atoms in CNT but to a lesser extent. We notice that the atoms, which are far away from the nucleoside (those of the edge-2), do not undergo any noticeable changes in both the cases. The partial charges of the central carbon atoms also alter significantly, particularly in case of uridine, which is reflected in larger standard deviations (partial charge of carbon changes from 0.002 to –0.060 for acting as H-bond acceptor). The other carbon atoms of CNT that are forming H-bonds with the O-H groups of uridine acquire Mulliken charges (in units of electron charge 'e') of -0.063 and -0.154 from 0.020 and -0.149, respectively. This signifies that the uridine can polarize CNT and has strong binding affinity with CNT than that of the thymidine, which is correlated with its interaction energy data also.

## 4. Conclusions

Using combination of all atom molecular dynamics simulations and ab-initio quantum mechanical calculations we report unzipping and wrapping of a small interfering RNA molecule on the carbon nanotube to study the binding mechanism of siRNA on the CNT surface. To the best of our knowledge this is the first theoretical demonstration of the unzipping and wrapping of siRNA on the CNT surface. In the process of understanding siRNA delivery mechanism to achieve RNAi by CNTs at a microscopic level, we attempt to study the thermodynamic and energetic properties of siRNA-CNT complex. The binding mechanism of siRNA on various CNT diameters has been studied. Our simulations show that a few base-pairs at both the strands of siRNA get unzipped and wrap around CNT surface with strong binding affinity. The binding energy increases with the CNT diameter due to van der Waals forces between siRNA aromatic rings and CNT surface. In order to wrap around CNT, siRNA has to be very flexible. Since siRNA molecule is double stranded with two sticky-ends whose persistence length is much larger than that of single stranded RNA, we may naively expect that it will be difficult to wrap. But interestingly, siRNA gets unzipped and eventually wraps around the CNT surface within a few nanoseconds for all the CNT diameters studied in this paper. The unzipping and subsequent binding processes were initiated and driven by van der Waals (dispersion) interaction between aromatic rings of siRNA and CNT surface, facilitated by two sticky-ends on both the strands. siRNA gains entropy on binding to the CNT surface due to unzipping of a few base-pairs. More surface area of CNT for large diameters enhances the interaction with siRNA and improves binding. An increase in broken WC H-bonds, RMSD and number of close contacts indicate large structural deformation of siRNA with respect to its starting crystal structure. The siRNA-CNT complex is very stable after the optimal binding to the CNT. The dsDNA of the same sequence adsorption on the same CNT show that dsDNA has very less unzipping and wrapping around the CNT in the observed simulation time scale of 70 ns. Considering the unzipping and wrapping process is guided by uridine-CNT interaction, we have carried out detail quantum chemical analysis of the two comparative systems, CNT-uridine and CNT-thymidine. Our quantum chemistry results indicate that CNT has better propensity to bind to uridine due to its additional O-H groups which form strong H-bond with the CNT. The H-bond formation requires charge accumulation towards some carbons of the CNT, which also takes place due to extended conjugation of the CNT. We also studied effect of salt concentration on the siRNA-CNT interaction. Our results suggest that at large NaCl concentration, the screening of ionic charges make it less efficient for binding compared to charge neutral case. However the unzipping and wrapping of siRNA happens very slowly in this case. The adsorbed siRNA can be delivered to virus infected cell via endocytosis to reduce the expression of specific unwanted genes to achieve RNAi effect. In RNAi therapy, siRNA delivery to the target cell involves siRNA unzipping into two single strands that is mediated by RISC loading complex [3]. We have shown the unwinding enhanced siRNA binding to carbon nanotube (CNT) and propose siRNA delivery to target cell for achieving RNAi without degradation. In a subsequent work we study the binding mechanism of siRNA on graphene for more efficient and sophisticated delivery method. The intrinsic toxicity effects caused by CNT to the cell functionality was shown to have less effect by proper functionalization of CNT [66-68]. Studies on cell penetrating membrane mechanism of siRNA-CNT complex and solubility of CNT after siRNA delivery are among future perspectives in this vast area.

## Acknowledgments

PKM and AKS thank DBT, India for the financial support. We thank Prof. M. Muthukumar for stimulating discussions. We acknowledge computational resource supported by the DST Centre for Mathematical Biology at IISc. We also thank Prof. S. Ramaswamy for allowing us to use his computational facility. MS thank UGC, India for senior research fellowship.

Refereces

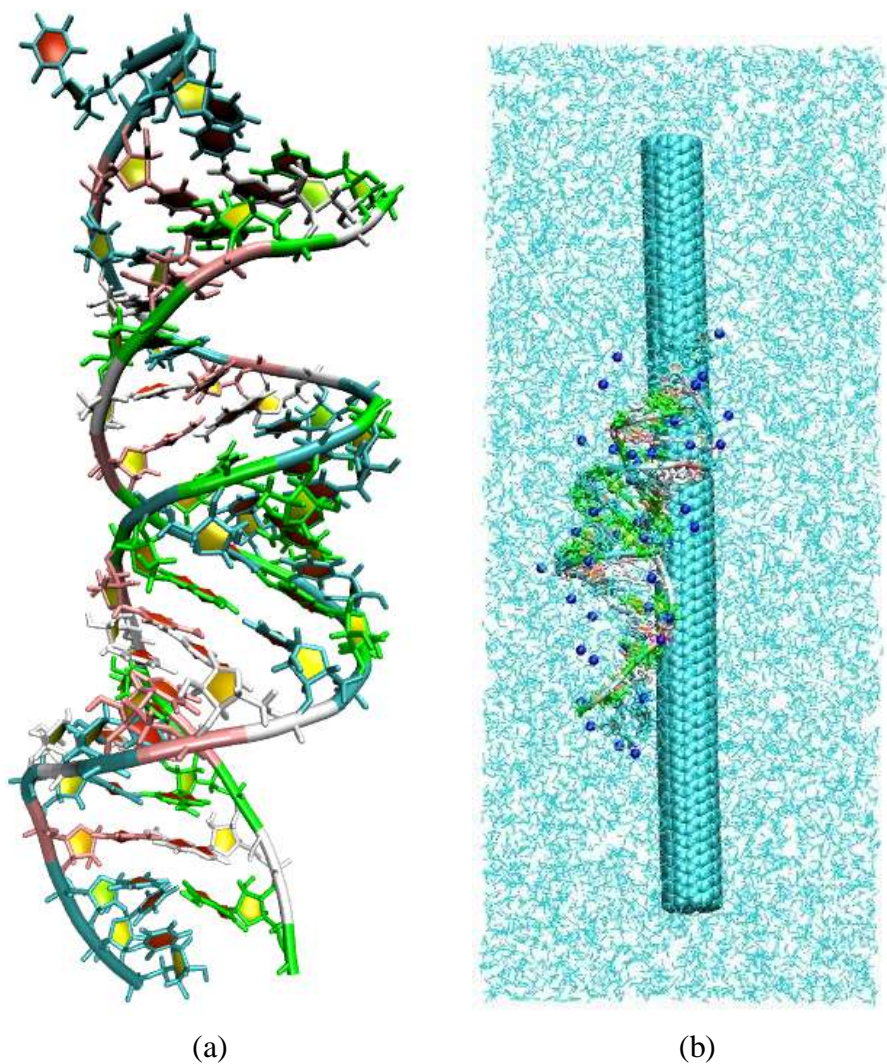

(a) (b)

Figure 1: (a) siRNA crystal structure (pdb code 1F8S) and (b) the initial simulation system setup where siRNA-(6, 6)CNT complex was solvated with water and neutralizing $Na^+$ counterions. These pictures were rendered using VMD software package [72].

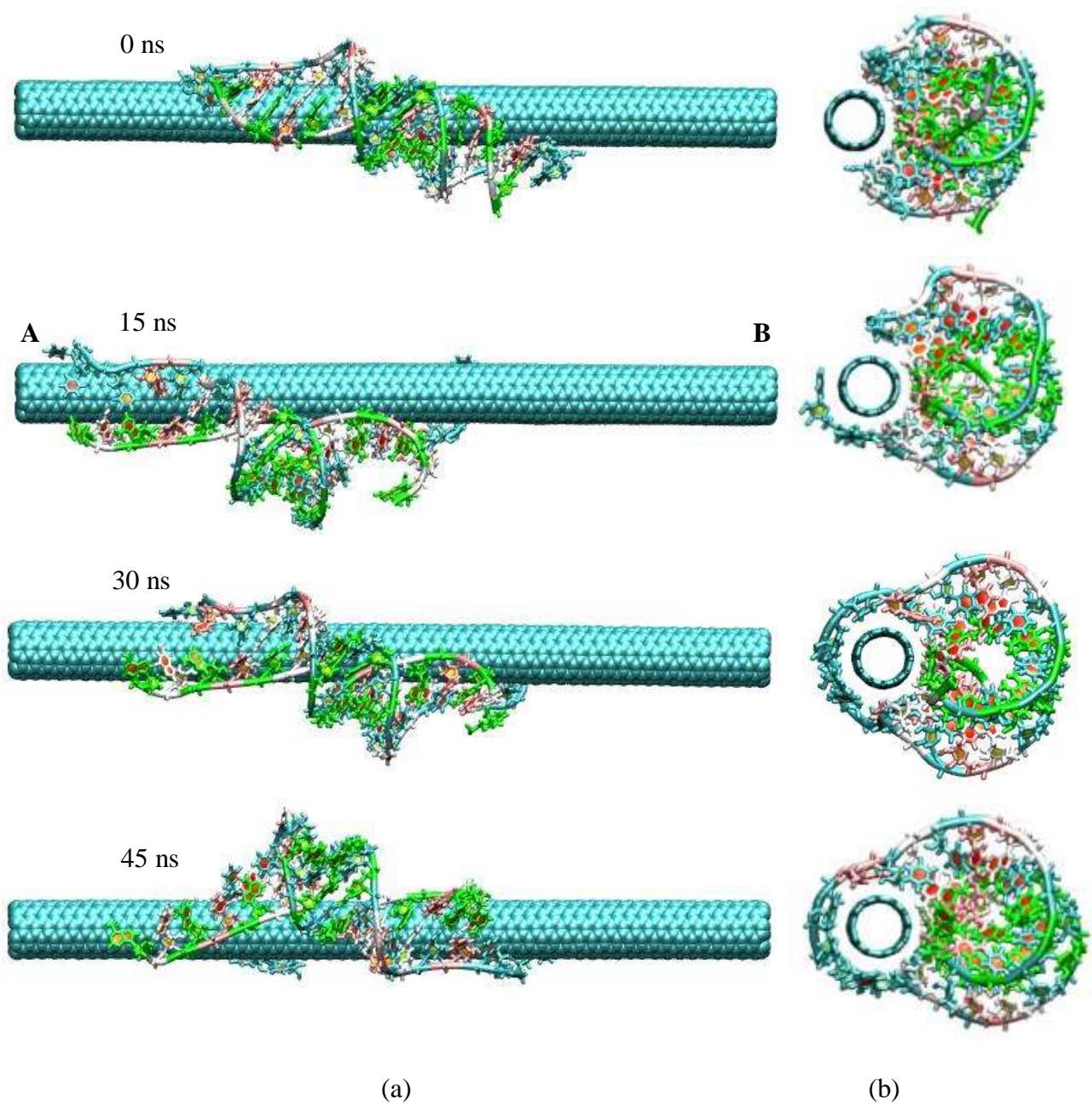

(a)          (b)

Figure 2: Snapshots of siRNA on (6, 6) CNT at 300 K at various instants of time during MD simulation in (a) horizontal and (b) vertical view with respect to CNT axis. Counterions and water molecules were not shown for clear visualization purpose. These snapshots were rendered using VMD software package [72].

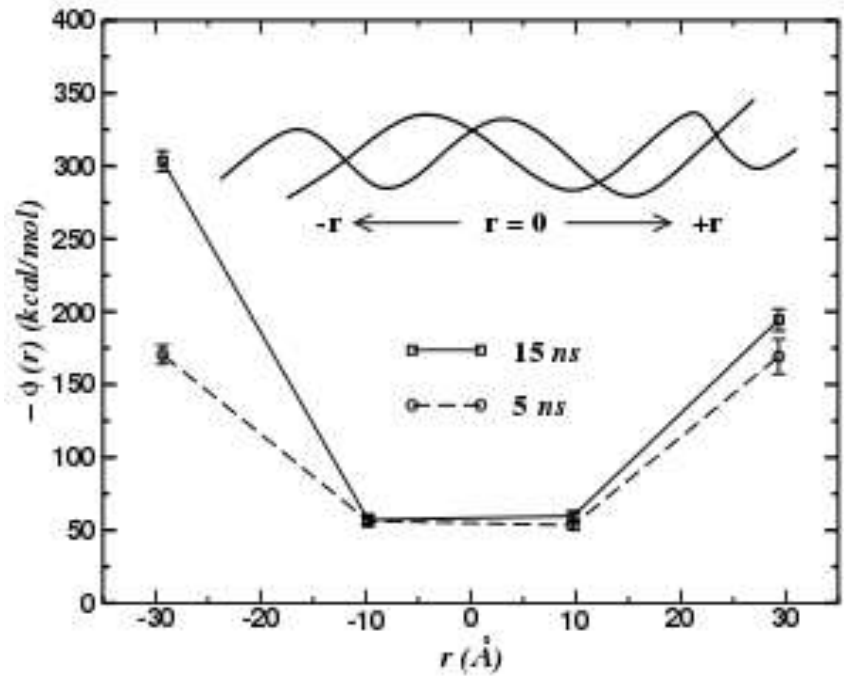

Figure 3: van der Waals interaction energy between siRNA and (6, 6) CNT at 300 K. See the text for details of the calculation. The siRNA strongly binds to CNT at 15 ns compared to 5 ns with an increment in van der Waals energy of about 133 kcal/mol or 225.4 $k_BT$.

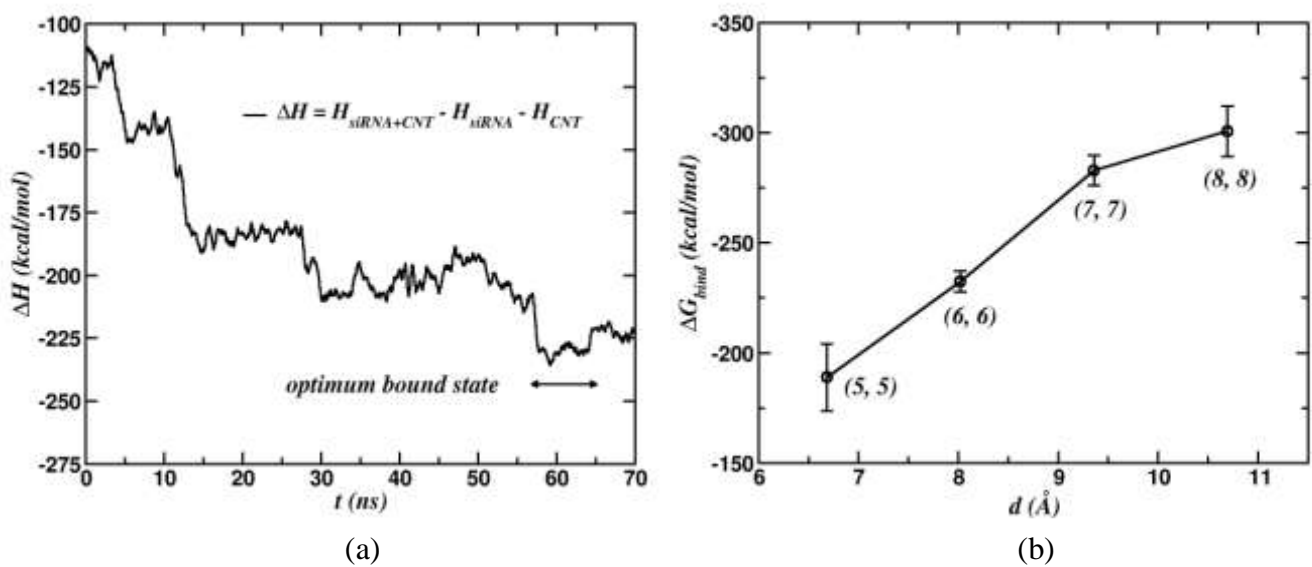

(a)  (b)

Figure 4: (a) The enthalpy contribution to the total binding free energy as a function of time for siRNA on (6, 6) CNT at 300 K. The arrow marked region shows the most optimum bound state with binding energy fluctuating within 5 % of its average in optimum bound state. For the entropy calculation we have started from this optimum bound state and simulated 10 consecutive sets of each 20 ps as discussed in text. (b) Binding free energy between siRNA and CNT in the most optimum bound state as a function of CNT diameter. The binding free energy that includes enthalpy and entropy contributions increases with diameter due to large CNT surface area available for siRNA to bind with CNT at 300 K.

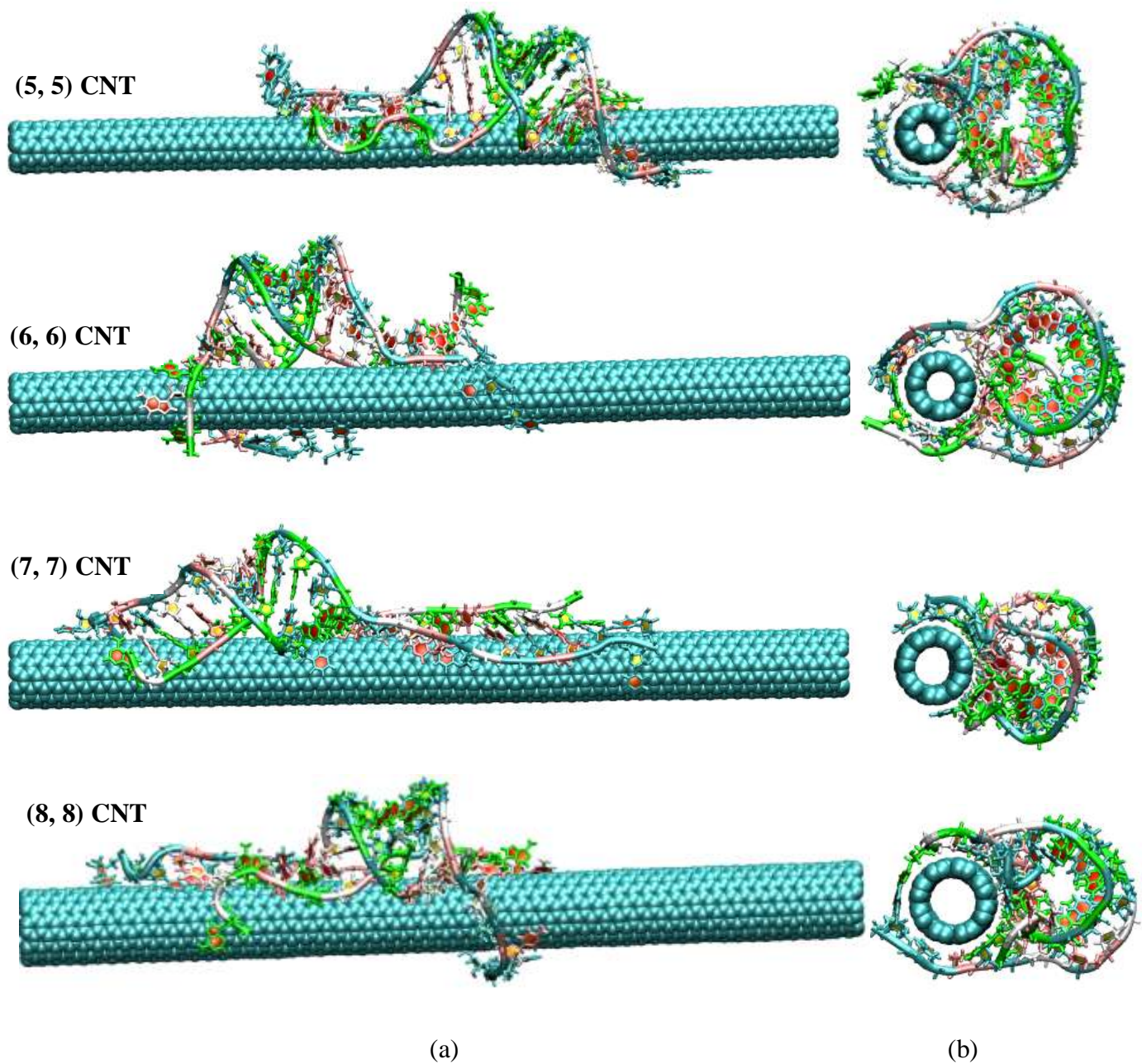

(a)  (b)

Figure 5: Snapshots of siRNA on (5, 5), (6, 6), (7, 7) and (8, 8) CNT at 300 K in the most optimum bound configuration in (a) horizontal and (b) vertical view with respect to CNT axis. Counterions and water molecules were not shown for clear visualization purpose.

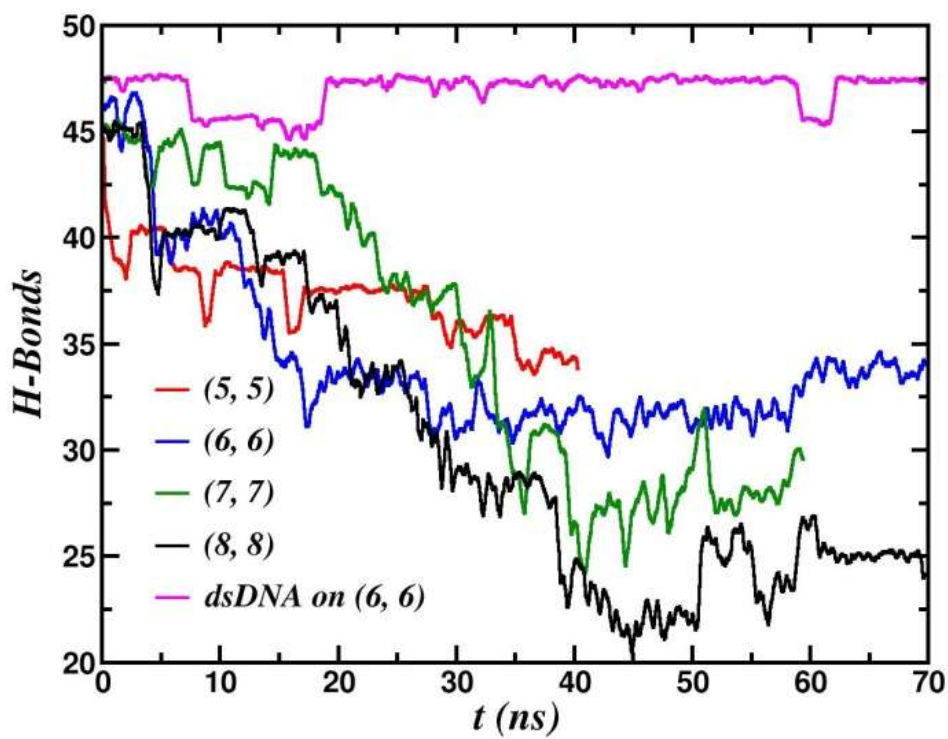

Figure 6: Time series of the number of intact Watson-Crick H-bonds in siRNA as a function of CNT diameter. In the optimum bound configuration, the number of intact H-bonds is decreasing with nanotube diameter implying a large adsorption to the nanotube with increasing diameter.

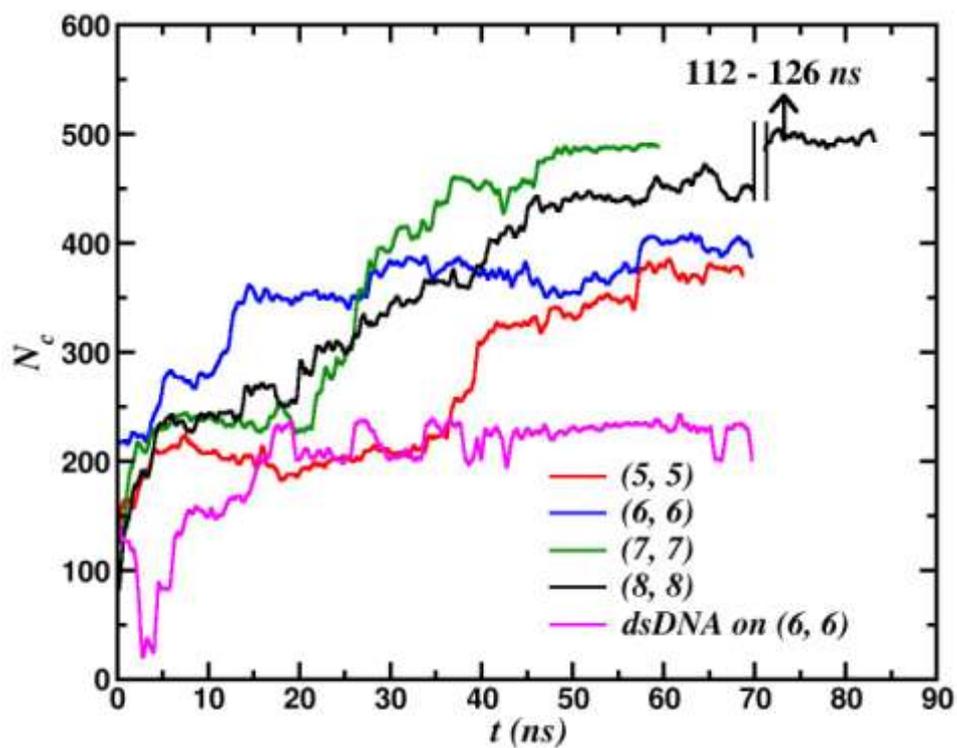

Figure 7: Number of close contacts ($N_c$) of siRNA to various CNT diameter within a cutoff of 5 Å from CNT surface. We simulate siRNA on (8, 8) CNT up to 126 ns to get the optimum bound configuration of the complex. In the plot we show $N_c$ from 112 to 126 ns for (8, 8) CNT with a line break after 70 ns.

Table 1: Summary of the simulation setup for 0 *mM*, 10 *mM* and 150 *mM* NaCl concentrations.

| c (mM) | CNT | siRNA | Box (Å$^3$) | Na$^+$ | Cl$^-$ | WAT | Total Atoms |
|---|---|---|---|---|---|---|---|
| 0 | (6, 6) 1440 | 1396 | 84×79×192 | 44 | 0 | 35382 | 109026 |
| 10 | (6, 6) 1440 | 1396 | 84×79×192 | 52 | 8 | 35365 | 108991 |
| 150 | (6, 6) 1440 | 1396 | 84×79×192 | 160 | 116 | 35148 | 108556 |
| 0 | (5, 5) 1200 | 1396 | 71×75×193 | 44 | 0 | 27810 | 86070 |
| 0 | (7, 7) 1680 | 1396 | 79×72×193 | 44 | 0 | 29810 | 92550 |
| 0 | (8, 8) 1920 | 1396 | 70×80×193 | 44 | 0 | 29063 | 90549 |
| 0 [a] | (6, 6) 1440 | 1398 | 78×67×192 | 42 | 0 | 26878 | 83514 |
| 0 [b] | (6, 6) 1440 | 1398 | 75×68×192 | 42 | 0 | 26472 | 82296 |

[a] dsDNA without sticky ends
[b] dsDNA with sticky ends

Table 2: Entropy per Na$^+$ ($S$/Na$^+$) and its gain ($\Delta S$/Na$^+$) when siRNA wrapped around CNT surface. Here, for calculating entropy gain we have used the entropy per Na$^+$ counterion which is 9.0 *cal/mol-K* or 37.96 *J/mol-K* when only siRNA is present.

| Entropy | 0 *mM* | 10 *mM* | 150 *mM* |
|---|---|---|---|
| $S$/Na$^+$ (cal/mol-K) | 9.0 | 9.8 | 12.53 |
| $\Delta S$/Na$^+$ (cal/mol-K) | 0.0 | 0.8 | 3.53 |

Table 3: Interaction energy data from the quantum chemical calculations for (6, 6) CNT.

| System | System Name | BSSE (*kcal/mol*) | Deformation Energy (*kcal/mol*) | $E_{Int.}$ (*kcal/mol*) |
|---|---|---|---|---|
| Fig. S3a(i) | CNT+Uracil | 3.27 | 1.49 | -9.64 |
| Fig. S3a(ii) | CNT+Thymine | 3.68 | 0.77 | -12.65 |
| Fig. S3a(iii) | CNT+Uridine | 5.36 | 1.79 | -18.72 |
| Fig. S3a(iv) | CNT+Thymidine | 5.94 | 1.00 | -16.25 |

Table 4: Average values of Mulliken charges of the seven types of atoms presented in (6, 6) CNT. Values given in parentheses are standard deviations.

| Systems | Edge-1 C of C-C group | Edge-1 C of C-H group | Edge-1 H | Edge-2 C of C-C group | Edge-2 C of C-H group | Edge-2 H | Middle Carbons |
|---|---|---|---|---|---|---|---|
| Isolated CNT | 0.020 (0.000) | -0.150 (0.000) | 0.124 (0.000) | 0.020 (0.000) | -0.150 (0.000) | 0.124 (0.000) | 0.002 (0.003) |
| CNT + uridine | 0.012 (0.027) | -0.149 (0.003) | 0.128 (0.002) | 0.020 (0.002) | -0.149 (0.001) | 0.126 (0.001) | 0.0005 (0.024) |
| CNT + thymidine | 0.018 (0.011) | -0.150 (0.002) | 0.125 (0.00) | 0.021 (0.005) | -0.150 (0.000) | 0.124 (0.001) | 0.003 (0.008) |